# STUDY OF THE DISSIPATIVE BINARY CHANNELS IN THE $^{107}$Ag + $^{58}$Ni REACTION AT 52 MeV/NUCLEON

J.C. Steckmeyer,[1,*] S. Aiello,[2] A. Anzalone,[3] G. Auger,[4] M. Bini,[5] B. Borderie,[6] R. Bougault,[1] B. Bouriquet,[4] A.M. Buta,[1] G. Cardella,[2] G. Casini,[5] S. Cavallaro,[3] J.L. Charvet,[7] A. Chbihi,[4] J. Colin,[1] D. Cussol,[1] R. Dayras,[7] N. De Cesare,[8] E. De Filippo,[2] A. Demeyer,[9] D. Doré,[7] D. Durand,[1] S. Feminò,[10] J.D. Frankland,[4] E. Galichet,[6,11] M. Geraci,[2] E. Gerlic,[9] L. Gingras,[12] F. Giustolisi,[2] P. Guazzoni,[13] D. Guinet,[9] B. Guiot,[4] S. Hudan,[4] G. Lanzalone,[2] G. Lanzanò,[2] P. Lautesse,[9] F. Lavaud,[6] J.F. Lecolley,[1] R. Legrain,[7,†] N. Le Neindre,[4] S. Lo Nigro,[2] O. Lopez,[1] F. Lo Piano,[3] L. Manduci,[1] L. Nalpas,[7] J. Normand,[1] A. Olmi,[5] A. Pagano,[2] M. Papa,[2] M. Pârlog,[14] G. Pasquali,[5] P. Pawłowski,[6] S. Piantelli,[5] S. Pirrone,[2] E. Plagnol,[6] G. Poggi,[5] G. Politi,[2] F. Porto,[3] M.F. Rivet,[6] F. Rizzo,[3] E. Rosato,[8] R. Roy,[12] S. Sambataro,[2,†] M. Samri,[12] M.L. Sperduto,[3] A.A. Stefanini,[5] C. Sutera,[2] G. Tăbăcaru,[14] B. Tamain,[1] L. Tassan-Got,[6] M. Vachon,[12] E. Vient,[1] M. Vigilante,[8] C. Volant,[7] J.P. Wieleczko,[4] and L. Zetta[13]

(CHIMERA - INDRA Collaboration)

[1]*LPC, IN2P3-CNRS, ISMRA et Université de Caen, F-14050 Caen-Cedex, France.*
[2]*INFN e Università di Catania, 95129 Catania, Italy.*
[3]*LNS e Università di Catania, 95123 Catania, Italy.*
[4]*GANIL, CEA et IN2P3-CNRS, B.P. 5027, 14076 Caen-Cedex 05, France.*
[5]*INFN e Università di Firenze, 50125 Firenze, Italy.*
[6]*Institut de Physique Nucléaire, IN2P3-CNRS, 91406 Orsay-Cedex, France.*
[7]*DAPNIA/SPhN, CEA/Saclay, 91191 Gif sur Yvette-Cedex, France.*
[8]*INFN e Dipartimento di Scienze Fisiche, Università di Napoli "Federico II", Napoli, Italy.*
[9]*IPN, IN2P3-CNRS et Université, 69622 Villeurbanne-Cedex, France.*
[10]*INFN e Università di Messina, 98100 Messina, Italy.*
[11]*Conservatoire National des Arts et Métiers, Paris, France.*
[12]*Laboratoire de Physique Nucléaire, Université Laval, Québec, Canada.*
[13]*INFN e Università di Milano, 20133 Milano, Italy.*
[14]*National Institut for Physics and Nuclear Engineering, Bucharest-Măgurele, Romania.*

## Abstract

The binary dissipative channels are characterized by the presence of two main fragments in the exit channel. They have been studied in the $^{107}$Ag + $^{58}$Ni reaction at 52 MeV/nucleon of bombarding energy. For that purpose a modified version of the Indra multidetector has been used in conjunction with a part of the Chimera multidetector. Preliminary results on the excitation energy and intrinsic angular momentum of the quasi-projectile are reported and compared to a dynamical calculation.

## 1. INTRODUCTION

At bombarding energies well below the Fermi energy regime, the binary dissipative collisions are characterized by the presence of two main fragments in the exit channel, the so-called quasi-projectile (QP) and quasi-target (QT) nuclei. These fragments have properties reminiscent of those of the projectile and target [1]. The kinetic energy spans the whole range down to the value of the Coulomb repulsion, most of the kinetic energy loss is converted into excitation energy and a fraction of the initial orbital angular momentum is transferred into intrinsic spin to the fragments.

---

*E-mail: steck@in2p3.fr
†deceased

In this energy range, most of the features of the binary dissipative collisions are well understood. The mean values as well as the second moments of the distributions of observables such as the mass, charge and energy, are described qualitatively by transport models based on the stochastic exchange of individual nucleons between the two reaction partners all along the interaction [2]. According to this model, an equal number of nucleons transferred on the average from one fragment to the other, and vice-versa, is expected to occur in the peripheral reactions associated with short interaction times. This leads to a final configuration in which no thermal equilibrium is reached (equal partition of the excitation energy). For more dissipative collisions, the larger interaction times modify the net flow of transferred nucleons in such a way that a thermal equilibrium tends to develop (equal temperature of the fragments as expected in the Fermi model). Although this scenario is commonly accepted, there are experimental results which show that non equilibrium effects are observed and that they are correlated with the net gain of nucleons experienced by the nuclei whatever the energy dissipation [3, 4].

When increasing the bombarding energy, the scenario depicted above becomes progressively more and more inadequate. Light charged particles (LCP) and intermediate mass fragments (IMF) are emitted in the early stages of the collision, at velocities ranging between those of the projectile and target, increasingly perturbing the pure binary character of the reaction. These particles have characteristics which allow to think that they arise in a dynamical process: the transverse kinetic energy spectra with respect to the beam are independent of the bombarding energy and the isotopic ratios differ from those of projectile and target [5, 6].

The mid-velocity component likely carries away an important fraction of the dissipated kinetic energy as well as of the orbital angular momentum. The study of the sharing of the excitation energy and angular momentum between the QP and QT at intermediate bombarding energies requires to explicitely consider the increasing deviations from the picture of a pure binary first-step of the reaction, which is usually assumed at low bombarding energy. That is one needs to determine the characteristics of QP and QT by excluding as much as possible the contribution of particles emitted at mid-velocity: one has to disentangle the particles statistically evaporated by the QP and QT from particles emitted in non-statistical processes. The difficulty is that these different contributions overlap both in time and space.

Studying the partition of the excitation energy and angular momentum allows to collect information on the equilibration times of these degrees of freedom. With the increase of the bombarding energy and the growth of the nucleon-nucleon collisions, the equilibration times are expected to become smaller. The experimental determination of the characteristic times brings constraints to the theoretical models and may help to specify some ingredients as the nucleon-nucleon cross section in the nuclear matter [7–10].

From an experimental point-of-view, studying the binary channels needs to perform the detection of the QP and QT residues. Due to the low recoil energy of the QT, typically a few tens of MeV, specific detection devices have to be used.

In this report preliminary results on the determination of the excitation energy and spin of the QP produced in binary channels in the $^{107}$Ag + $^{58}$Ni reaction at 52 MeV/nucleon are presented. They are also compared to the predictions of the dynamical model of Ref. [2].

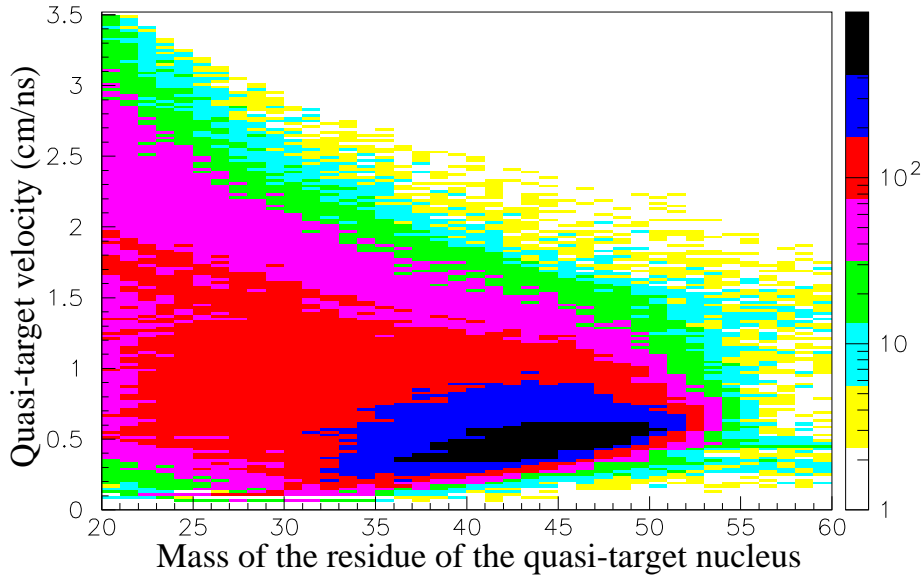

FIG. 1: For the QT residue: correlation between the velocity in the laboratory system and the mass.

## 2. EXPERIMENTAL SETUP

Due to the low kinetic energy of the QT, the Indra setup [11, 12] was modified: standard modules composed of an ionisation chamber, silicon detectors and cesium iodide scintillators have been removed and replaced by specific silicon detectors measuring the energy and the time of flight of ions with a flight path ranging from 50 to 125 cm depending on the detection angle. They were 10 detectors located in the horizontal plane at angles between 3° and 87°, each of them having four vertical strips.

At the bombarding energy of 52 MeV/nucleon, the QP is expected to be very forward peaked. To improve the angular coverage and the identification of the QP, the first ring of Indra made of plastic scintillators and covering the 2-3° angular range was replaced by the first wheel of the Chimera detector [13] made of two rings, each of them composed of sixteen Si-CsI telescopes settled down in the 1-3° angular range. The measurement of the time of flight of ions hitting the silicon detectors allows for an average mass determination.

## 3. EXPERIMENTAL DATA

The experimental events were triggered by the detection of a charged product in one of the ten silicon detectors devoted to the detection of the QT. Most of the detected products are LCP and IMF. Therefore the binary channels have been selected by requiring the detection, in one of the Si detectors, of a nucleus with a mass larger than 20 u. Additional selections required the total collected charge and linear momentum to be larger than 90% of the incident values, respectively. In Fig. 1 is shown for the QT residue, the correlation between the velocity in the laboratory system and the detected mass. This correlation is integrated over the whole angular range of the silicon detectors devoted to the detection of the QT nuclei. The angular distribution of these nuclei is broad, ranging from very forward angles to large angles. The yield of the QT nuclei is slightly increasing with the increase of the detection angle. A large fraction of the QT have velocities smaller than 1 cm/ns, i.e. kinetic energies lower than 20-30 MeV.

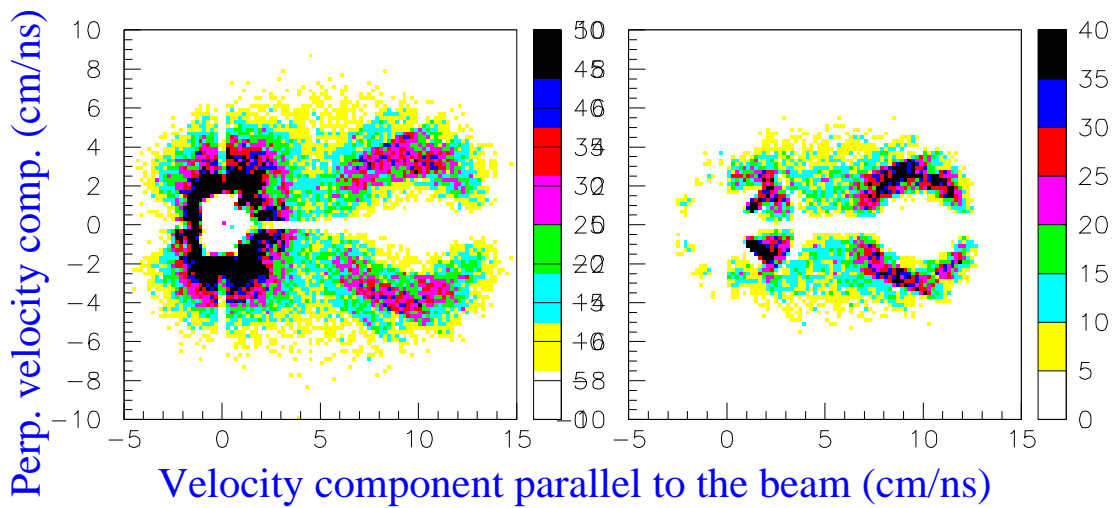

FIG. 2: Velocity plots: velocity component perpendicular to the beam as a function of the velocity component parallel to the beam, for protons (left) and alpha particles (right).

In Fig. 2 velocity plots for protons and alpha particles are shown. They correspond to binary events measured with a total multiplicity of charged products between 9 and 12. The $V_\perp$-$V_\parallel$ plane is the reaction plane defined by the beam direction and the QP recoil velocity. By construction, the QP direction is assigned to a negative transverse velocity in Fig. 2. Two sources are clearly recognized, the target source corresponding to particles lying on a circle approximately centered on the zero velocity and the projectile source associated with a velocity of the order of 9-10 cm/ns. Particles are present between the two sources: they contribute to the mid-velocity component. At backward angles higher identification thresholds for alpha particles than for protons are clearly visible in Fig. 2. More alpha particles are emitted on the other side of the QP with respect to the beam, as already seen in the $^{129}$Xe+$^{nat}$Sn at 50 MeV/nucleon [14].

From the velocity correlations displayed in Fig. 2, it is stated that the particles emitted with a velocity higher than the QP velocity do come from the projectile, otherwise the Coulomb correlation with no particle inside the ring would not be observed.

### 4. ANALYSIS

By selecting particles which are unambiguously emitted by the projectile source, comparison can be performed between the data and predictions of models describing the disintegration properties of excited projectiles. The Gemini [15] code has been chosen as it is widely used in the community and it will allow for testing the assumption of a statistical evaporation from the projectile.

In Fig. 3 is shown the evolution of the atomic number of the residue of the source as a function of the excitation energy and for different values of the angular momentum. The calculations have been performed for a $^{107}$Ag nucleus. As seen, the value of the atomic number of the residue appears to be independent of the spin. Deviations are only seen for low excitation energies and huge angular momenta. From the correlation shown in Fig. 3 it is possible to associate, on the average, an excitation energy to a given charge of the QP residue.

Looking at the LCP multiplicities, an increase is observed with the increase of the excitation energy.

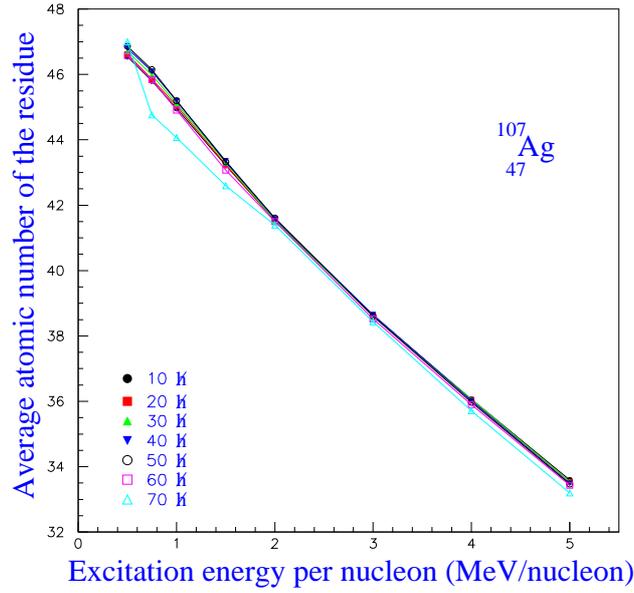

FIG. 3: Atomic number of the residue as a function of the excitation energy per nucleon and for different values of the angular momentum J.

However the proton multiplicity decreases with the increase of the angular momentum, while the alpha multiplicity increases with the increase of the angular momentum. The opposite evolution of the proton and alpha multiplicities allows for a determination of both the excitation energy and the spin of the source as far as the experimental multiplicities agree with the range of multiplicities predicted by Gemini. To a given couple ($M_p$, $M_\alpha$) can be associated a couple ($E^*/A$, J) where $E^*/A$ is the excitation energy per nucleon and J the spin [16].

For the sake of comparison, the data have to be sorted as a function of a parameter related to the energy dissipation occurring during the collision. In this work, the QP recoil velocity has been chosen as it can be connected to the total kinetic energy loss (TKEL).

## 5.  RESULTS

In Fig. 4 is shown the J-$E^*$/A correlation deduced either from the $M_p$-$M_\alpha$ multiplicities (open circles) or from the $M_H$-$M_{He}$ multiplicities (open squares). The higher the excitation energy, the higher the intrinsic angular momentum. The highest values of the spin range between 70 and 80 $\hbar$. The angular momentum at which the fission barrier of a $^{107}$Ag nucleus vanishes is 74 $\hbar$ [17]. Typical uncertainties on the estimation of the spin values are $\approx$ 5 $\hbar$. For the excitation energy an uncertainty of 0,35 MeV/nucleon is associated to an uncertainty of one unit on the atomic number of the QP residue.

The excitation energy per nucleon of the QP as a function of the parallel velocity in the laboratory system is displayed in Fig. 5. The excitation energy (filled triangles) has been obtained from the correlation shown in Fig. 3. The increase of the QP excitation energy is clearly correlated to the decrease of the QP velocity: the slowing down of the velocity is linked to an increase of the energy dissipation and then of the intrinsic excitation energy. In Fig. 5 the excitation energy (open circles and squares) deduced from the J-$E^*$/A correlation obtained from the LCP multiplicities (see above) is also plotted.

The slight difference observed in the estimation of the excitation energy using either the average atomic number of the residue or the LCP multiplicities can be attributed to the IMF emission. Indeed the experimental value of the QP residue is the result of a decay chain in which LCP and IMF are evaporated. In fact an emission of IMF is experimentally observed, although it is rather weak. In the calculation this contribution is not taken into account as the Gemini code does not reproduce correctly the multiplicity of light clusters. As a consequence the excitation energy deduced from the LCP multiplicities alone is underestimated.

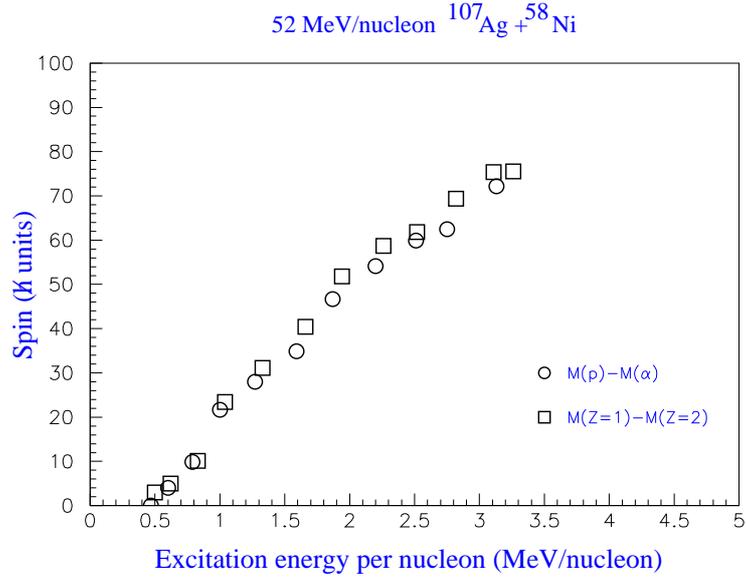

FIG. 4: The J - E*/A correlation deduced from the $M_p$-$M_\alpha$ multiplicities (open circles) and from the $M_H$-$M_{He}$ multiplicities (see text).

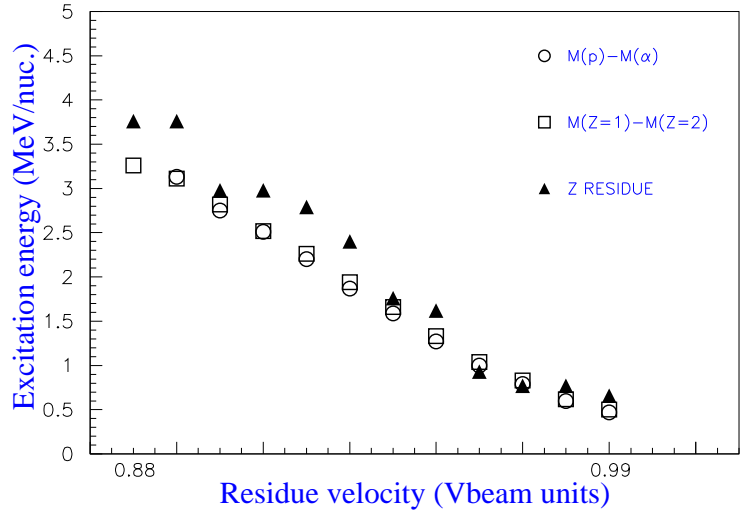

FIG. 5: The excitation energy per nucleon of the QP as a function of its velocity. Open symbols are derived from the coupled information of $M_p$-$M_\alpha$ or $M_H$-$M_{He}$ multiplicities, while filled triangles are deduced from the correlation shown in Fig. 3.

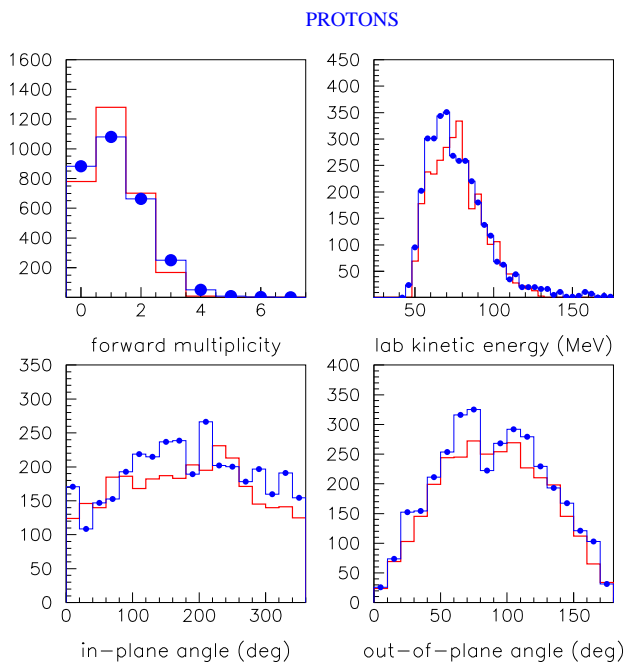

FIG. 6: Filled symbols and thin lines represent the data and the thick lines the predictions of Gemini model [15]. These results are dealing with the protons. Upper left corner: proton multiplicity in the forward hemisphere of the source, upper right corner: laboratory kinetic energy spectrum, lower left corner: in-plane angular distribution and lower right corner: out-of-plane angular distribution.

## 6. COMPARISON BETWEEN THE DATA AND THE GEMINI SIMULATION

In order to learn more on the properties of QP in binary collisions, detailed comparisons between the data and the simulation have to be done. In the simulation, the projectile source has been given a velocity distribution in such a way that the simulated velocity distribution of the QP residue describes the measured one. Once the source velocity is determined, the properties of LCP are derived. They are displayed for protons in Fig. 6. The filled circles and thin lines represent the data, the thick lines the results of the simulation. These results deal with the velocity bin associated with an excitation energy of $\approx 2.2$ MeV/nucleon and a spin value of $\approx 60\,\hbar$ (cf. Fig. 4).

The forward multiplicity, the kinetic energy spectrum of the forward emitted protons and their in-plane and out-of-plane angular distributions in the frame of the detected QP are shown respectively. The experimental reaction plane is defined by the beam direction and the QP recoil velocity. The out-of-plane angle is the angle between the spin axis, perpendicular to the reaction plane, and the velocity of the emitted particle, while the in-plane angle is the angle between the QP recoil velocity and the projection of the LCP velocity in the reaction plane. In Fig. 6 the in-plane angular distribution has a maximum value at 180°: the particle and residue are emitted back to back. The out-of-plane angle has a maximum at 90°: the protons are mainly emitted in the reaction plane.

The simulation performed with the statistical Gemini code describes the data in a quantitative way. The experimental trends are well reproduced. The agreement is even better for the alpha particles (not shown here). In particular their angular distributions are narrower, both in the experiment and in the simulation. From that observation it can be argued that a standard statistical calculation describes the de-excitation properties of QP produced in binary collisions.

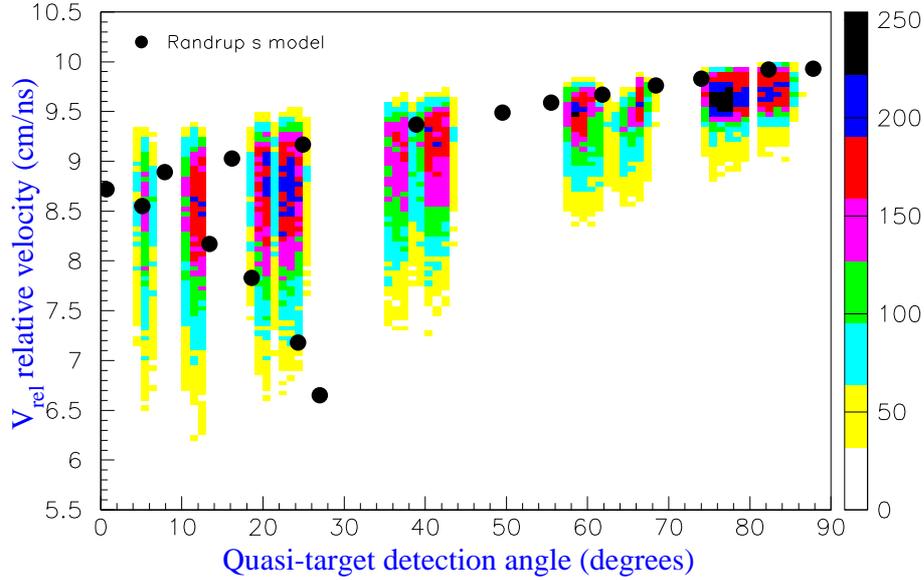

FIG. 7: The relative velocity between the QP and QT residues as a function of the QT angle in the laboratory system. Black points are the predictions of the dynamical model [2].

## 7. COMPARISON WITH DYNAMICAL CALCULATIONS

### 7.1. Kinematics

Many aspects of experimental data collected at low bombarding energies are understood in the framework of the stochastic nucleon exchange model (NEM) [2]. The mean values of mass and charge distributions as well as their second moments are qualitatively reproduced. The energy dissipation is well accounted for via the description of kinematic variables, in particular the deflection angle. Other degrees of freedom as the angular momentum are also taken into account.

J. Randrup developed a general theory of transport induced by nucleon transfer [18]: the exchange of individual nucleons produces a transport of mass, charge, energy, linear and angular momentum. The energy dissipation relies on the one-body proximity friction formalism [19]. The dinuclear composite is depicted as two spherical nuclei approximated by two Fermi gases and joined by a cylindrical neck.

In Fig. 7 is displayed the correlation between the relative velocity of the two partners and the deflection angle of the QT residue in the laboratory frame. This correlation looks like a Wilczynski plot usually plotted in the low energy domain showing the correlation between the total kinetic energy and the deflection angle in the center of mass system. In Fig. 7, the peripheral reactions are associated with the highest values of relative velocities ($V_{rel}$) and deflection angles ($\Theta_{QT}$), while small values of $V_{rel}$ and $\Theta_{QT}$ correspond to more dissipative collisions.

The results of dynamical calculations performed with the NEM model are plotted as black points in Fig. 7. The calculated $V_{rel}$ velocity is obtained from the difference between the velocities of the primary QP and QT nuclei left just after the collision: there is no de-excitation of the hot primary nuclei in the model (nor any pre-equilibrium or mid-velocity emission). On the other hand, the experimental $V_{rel}$ velocity is built from the velocities of cold residues. Even if it is usually assumed that the evaporation process does not modify on average the emission direction of the residues, this difference has to be kept in mind in the confrontation of the data and calculations.

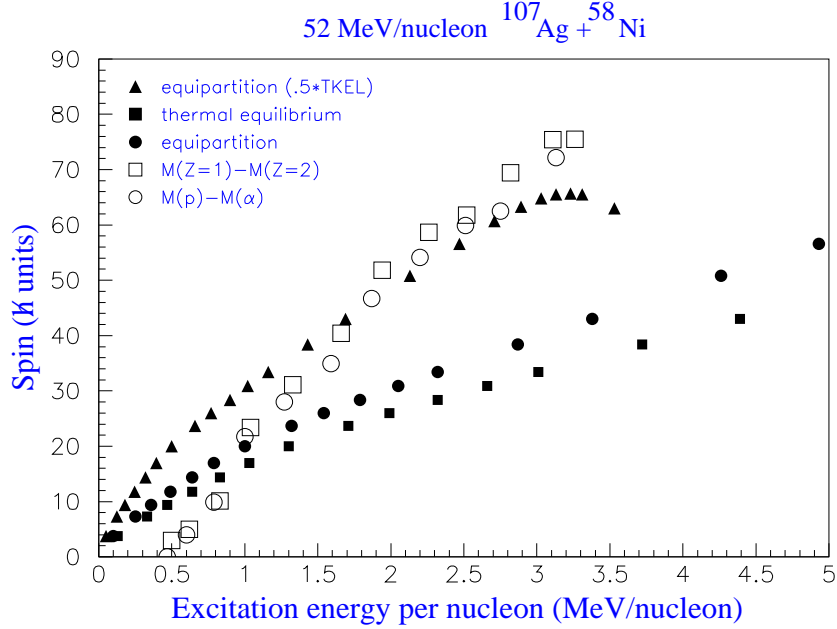

FIG. 8: Correlation between the intrinsic spin and the excitation energy per nucleon for the QP nucleus. The data are represented by the open symbols and the predictions of the dynamical model of Ref. [2] are given by the filled symbols (see text for explanation).

The calculated dynamical trajectory starts at an angular momentum J of 740 $\hbar$. With the decrease of J (and of the impact parameter) the dissipation sets in. In particular the deflection angles of both nuclei vanish at a value of J = 605 $\hbar$ (approximately 440 MeV of total kinetic energy loss). At that time the trajectories of the two nuclei are bent towards the negative angles. The last point of the calculated trajectory is associated with an angular momentum of 540 $\hbar$ and a TKEL value of 1050 MeV. As mentioned above, more detailed reconstructions are needed in order to reach quantitative conclusions. However it can be seen that the results of the calculations are in a reasonable agreement with the kinematical features displayed by the data.

### 7.2. Angular momentum transfer

The treatment of the angular momentum observables is explicitly accounted for in the NEM model [2]. Further improvements were even added to the model [20].

In Fig. 8 the data already shown in Fig. 4 are compared with the predictions of the dynamical calculations. As the spin value is reported as a function of the excitation energy of the quasi-projectile, the TKEL predicted by the NEM model has been converted in an excitation energy scale employing two hypotheses: an equipartition of the excitation energy between the fragments (filled dots in Fig. 8) or a sharing according the mass ratio of the nuclei (thermal equilibrium) (filled squares in Fig. 8). As seen the calculations and data are not in agreement. For excitation energies larger than 2 MeV/nucleon, the calculated spins are too low by a factor of two, while for the low excitation energies the spin value is overestimated. This last discrepancy is attributed to the description of the neck degree of freedom and would call for a better treatment of the nucleon transfer at low TKEL as noticed by the authors of Ref. [20]. At low bombarding energies and high TKEL, the same authors observe the same behaviour as the one displayed in Fig. 8 for the higher excitation energies. They associate this discrepancy to the absence of the treatment of the dynamical fluctuations which prevents the highest TKEL values to receive contributions from a wide range of impact parameters.

In case of the $^{107}$Ag + $^{58}$Ni reaction at 52 MeV/nucleon, we would rather invoke an inadequacy between the data and the calculations about the pure binary character of the collision. Assuming that only half of the TKEL is converted into excitation energy leads to the filled triangles in Fig. 8. And this would be more in agreement with the excitation energy deduced for the QP, as in the most dissipative collisions studied here a total excitation energy would approach $\approx$ 600 MeV (assuming an energy sharing in the mass ratio) while the estimation of the NEM model is 1050 MeV (see above).

Although depicting the dynamics of the collision in a reasonable way, the NEM model fails in reproducing the angular momentum transferred to the projectile. The confrontation has to be done in a more detailed way, but the fact that the model does not account for a non-equilibrium emission likely prevents its use at these intermediate bombarding energies, and more sophisticated models as the one of Ref. [21] have to be called for.

## 8. CONCLUSION

The binary channels have been studied in the $^{107}$Ag + $^{58}$Ni reaction at 52 MeV/nucleon. The data exhibit an increase of the excitation energy of the QP when the velocity decreases. A nice correlation is observed between the intrinsic spin and the excitation energy. The experimental trends are well reproduced by standard statistical calculations. The highest excitation energy reaches 3-4 MeV/nucleon while the spin goes up to 70-80 $\hbar$.

A preliminary attempt to compare the data with the predictions of a dynamical transport model shows a qualitative agreement on the reaction dynamics, while the angular momentum transferred to the quasi-projectile is not reproduced. It is mentioned that a treatment of the mid-velocity emission should be likely incorporated in order to perform more conclusive confrontation with the data.